\documentclass[12pt]{book}

\usepackage[dvips]{graphicx,color}
\usepackage{makeidx,tsukuba}
\makeauthorindex
\makeindex

\begin{document}

\BookTitle{\itshape The 28th International Cosmic Ray Conference}
\CopyRight{\copyright 2003 by Universal Academy Press, Inc.}
%\tableofcontents
\pagenumbering{arabic}

%%%%%%%%%%% The first letter of each word should be capital letter.
\chapter{Search for TeV annihilation radiation from supersymmetric dark matter in nearby galaxies}

\author{%
%
% You can include as many co-authors as you wish, unless
% the title/author information fits within 1 page.
%
V.V.~Vassiliev$^{1,2}$, I.H.~Bond, P.J.~Boyle, S.M.~Bradbury, J.H.~Buckley,
D.~Carter-Lewis, O.~Celik, W.~Cui, M.~Daniel, M.~D'Vali,
I.de~la~Calle~Perez, C.~Duke, A.~Falcone, D.J.~Fegan, S.J.~Fegan,
J.P.~Finley, L.F.~Fortson, J.~Gaidos, S.~Gammell, K.~Gibbs,
G.H.~Gillanders, J.~Grube, J.~Hall, T.A.~Hall, D.~Hanna, A.M.~Hillas,
J.~Holder, D.~Horan, A.~Jarvis, M.~Jordan, G.E.~Kenny, M.~Kertzman,
D.~Kieda, J.~Kildea, J.~Knapp, K.~Kosack, H.~Krawczynski, F.~Krennrich,
M.J.~Lang, S.~LeBohec, E.~Linton, J.~Lloyd-Evans, A.~Milovanovic,
P.~Moriarty, D.~Muller, T.~Nagai, S.~Nolan, R.A.~Ong, R.~Pallassini,
D.~Petry, B.~Power-Mooney, J.~Quinn, M.~Quinn, K.~Ragan, P.~Rebillot,
P.T.~Reynolds, H.J.~Rose, M.~Schroedter, G.~Sembroski, S.P.~Swordy,
A.~Syson, S.P.~Wakely, G.~Walker, T.C.~Weekes, J.~Zweerink$^2$, 
and B.C. Bromley$^1$\\
{\it
(1) University of Utah, 115S 1400E, rm 201, Salt Lake City, UT 84112, USA}\\
{\it
(2)The VERITAS Collaboration-- see S.P.Wakely's paper} ``The VERITAS
Prototype'' {\it from these proceedings for affiliations}
}%% end of author

\section*{Abstract}

During the 2002-2003 observing season the Whipple 10m imaging atmospheric 
\v{C}herenkov telescope was used to search for dark matter annihilation 
radiation in four nearby galaxies: M32, M33, Draco, and Ursa Minor. 
Scientific motivations for this choice of targets are discussed as well as
accumulated exposure. The analysis results are to be reported in 
the conference presentation.

\section{Introduction}

The lightest supersymmetric particle, the neutralino, with a mass 
in the range $50$ GeV - $5$ TeV, is a plausible candidate for non-baryonic 
cold dark matter (CDM) [12, 16], which can be detected indirectly via its  
annihilation products [3, 32]. The annihilation rate is proportional to 
the square of the neutralino density integrated along the line of sight, 
suggesting strong enhancement in the direction of dark matter clumps. 
The increasing dark matter density profile toward the center of Milky 
Way (MW) galaxy and its proximity made the Galactic Center a natural choice 
for such searches [4, 5, 19]. The CDM annihilation flux from even nearby 
extragalactic objects, such as dwarf galaxies located only ten times 
father, would have to be suppressed by a factor of hundred. However, 
due to the squared neutralino density dependence, the distance penalty can 
be overcome by the equal increase of DM density in the cores of MW satellite 
and local group galaxies. Although little is known about the distribution 
of DM in the central $1$ - $10$ pc of Galactic Nuclei (GN), which is the 
most relevant region for generating an annihilation signal, the current 
paradigm suggests that it is a strong function of the merger history of a 
galaxy [25] and its long-term evolution that could be affected by the 
presence and growth of central black holes (BHs), e.g. [13]. 
Because of this, D. Merritt argues that MW was perhaps unfortunate, 
and a DM spike in its center is unlikely [26]. 
On the other hand, if stellar, and stellar velocity distributions 
can be used as a guide to the distribution of dark matter at the 
very centers of galaxies, we suggest the following arguments to justify 
the choice of several extragalactic sources as plausible candidates 
for observation of DM annihilation. 

\section{M32}

M32 is the closest compact elliptical galaxy believed to be formed 
in a rare cosmological event from a low luminosity spiral galaxy when 
it plunged into the central region of M31 and most of the outer 
stellar component was tidally stripped [2]. Stellar kinematic 
data as well as gas-dynamical studies strongly support the presence 
of a single supermassive compact object, $\sim 3.6 \times 10^6 M_o$ 
in the center of M32 [15]. The core of M32 has a relatively homogeneous 
stellar population that can be modeled as being coeval and of 
intermediate age ($\sim 4$ Gyr) [7, 8]. Lauer et al. estimate M32 
core relaxation time scale as $\sim 2-3$ Gyr [22]. These data suggest that 
the nucleus of M32 was unlikely to have undergone a recent merging event 
when a massive black hole binary could have depleted the central stellar 
density by evacuating stars and destroying potential dark matter cusps 
in the galaxy core [27]. The data also imply that enough time 
has passed for collisional two-body relaxation of stellar population
around a black hole to form a cusp in the Stellar Density 
Profile (SDP), $\sim 1/r^{\alpha}$, with  $\alpha=3/2-7/4$ correspondent
to Bahcall-Wolf solution [1]. Optical and infrared data indicate 
SDP compatible with $\alpha$ in the range 
$1.4$ to $1.9$ and still rising at the resolution limit of 
$0.07$ pc. The exact spectral index and the possibility of its
break at a distance $\sim 1$ pc from the GN
center is being debated [7, 22]. The stellar density 
at the very center of M32's nucleus likely exceeds $10^7 M_o$ per 
pc$^3$, which is the highest known to us in nearby systems [22]. 

After initial violent relaxation during large scale structure 
formation non-interacting dark matter has no means of 
self-evolution on subgalactic scales because it cannot 
cool down without gravitational coupling to baryonic matter.
This is true at least for spherically symmetrical systems.
Thus it is likely that during formation of the galactic 
core the DM Density Profile (DMDP) should follow SDP 
with a characteristic relaxation time comparable to 
stellar relaxation that, in turn, is decreased by the heat 
transfer associated with the DM component in the nucleus.  
In addition, the slow growth of the central BH should 
cause adiabatic compression of DM with the adiabatic invariant 
$r M_{BH}=$const [20]. If almost all BH mass is acquired 
by slow growth, the amount of compression could 
be substantial, resulting in the appearance of a spike in the 
DMDP with $\alpha>2.25$ [13]. If, however, galactic 
mergers have been dominantly responsible for BH growth 
this may not only destroy potentially created DM spikes, it 
will also drastically suppress the amount of adiabatic DM compression 
around the BH. Probably both the effect of adiabatic BH growth and 
the evolution of DMDP driven by baryonic stellar and gas 
components in the cores of the galaxies take place at the same time 
and cannot be completely disentangled [23]. For the case of M32 both   
scenarios of DM evolution seem to favor a cusp in DMDP
with $\alpha \geq 1.5$.    

\section{M33}

The neutralino annihilation flux from a ``cuspy'' DMDP ($\alpha>1.5$) 
is formally divergent at small scales. The physically justified 
minimal radius is set by the equality between the annihilation rate 
and the rate of supply of neutralinos into annihilation 
region. The latter is related to the BH growth 
rate and/or relaxation rate of baryonic \& DM components in 
GN. The general effect of the central BH is to increase the stellar velocity 
dispersion within the radius of its gravitational influence [30], and 
consequently increase GN relaxation time, limiting neutralino 
annihilation flux dynamically. From this point of view a galaxy with a small 
BH and very rapid GN evolution scale, such as M33,  may be 
preferable for observations.

M33 is a normal low-luminosity dark matter dominated bulgeless spiral 
galaxy with dark halo $\sim 5.1 \times 10^{11} M_o$ [6]. The mass of 
the BH in its center is less than $\sim 1.5 \times 10^3 M_o$ [11, 24], 
and the stellar population in its nucleus can be modeled by two bursts 
of star formation $\sim 2$ and $\sim 0.5$ Gyr ago [29]. The nucleus of M33 
hosts the most luminous steady X-ray source in the Local Group that is also 
associated with a radio source and reminiscent of the galactic microquasar 
GRS 1915+105 [9]. Small ($\sim$\%$10$) but significant X-ray spectrum 
and time variability have also been reported [21]. M33 is remarkable 
for its very small galaxy nucleus relaxation time $\sim 3$ Myr due to
very dense, $>2 \times 10^6 M_o$ per pc$^3$, stellar core ($\sim 1$ pc)
and extremely low velocity dispersion [22].

\section{Draco and Ursa Minor dwarf galaxies}

Dark matter rich MW satellite Dwarf Galaxies (DGs) are favored for 
observations due to their proximity, very low baryonic content, 
and/or possibility of self-interacting DM. Within non-interacting 
DM scenarios the cores of these galaxies are ``frozen'' due to 
particularly low stellar and gas densities and consequently slow
evolution rates. Their typical core relaxation times, 
$\sim 5 \times 10^{2}$ Gyr, exceed Hubble time. 
Thus, the DGs could not have evolutionarily built any 
sub-parsec cusp or spike-like structure in their DMDP
for efficient annihilation of neutralinos, and it is also 
difficult to imagine any DM evolutionary scenario
on the scale of $\sim 1$ pc, a typical separation 
between stars in the nuclei of DGs. 
Tyler argues that these perhaps might be the exact 
conditions which would preserve intact any sub-parsec 
DM structures from destruction by galaxy merger events or by 
particularly high abundance of stars in GN [31]. 
The initial perturbations in DMDP with the scales 
smaller than the Jeans instability scale could have formed during 
violent relaxation and survived without trapping a substantial amount 
of baryons [18, 28]. These invisible and very slowly evolving 
structures could have been gravitationally trapped in the MW 
halo or in the centers of dwarf galaxies. If so, the quantitative 
prediction of the annihilation flux from DM cusps will strongly depend 
on the cusp evolution mechanism because it will determine the maximal 
annihilation rate. This scenario becomes particularly 
interesting if DM is self-interacting indeed [14]. 
Among many other puzzling observations of dwarf galaxies there 
are indications of ``clumpy'' distribution of matter. For example,
Ursa Minor dwarf has distinct stellar lumps within $\sim 10'$ circle [17].
The study of stellar proper motion suggests that the lump crossing time is
$\sim 5$ Myr and six individual lumps shouldn't exist any longer than 
this [10]. Is it a coincidence to observe six of them at the same time?

\section{Data}

Table 1 shows the range of zenith angles at which all proposed 
targets were observed and the total exposure so far accumulated 
by the VERITAS collaboration for each source.

\begin{table}[t]
 \caption{Accumulated exposure}
\begin{center}
\begin{tabular}{l|ccccc}
\hline
Source               & ON (hrs) & OFF (hrs)     & Zenith angle range    \\
\hline
M32                  & 10 & 4.5     & $10^o-40^o$    \\
M33                  & 13 & 4.0     & $1^o-40^o$     \\
Draco dwarf          & 4.5 & 3.0     & $27^o-35^o$   \\
Ursa Minor dwarf     & 5.5 & 4.0     &  $35^o-39^o$  \\
\hline
\end{tabular}
\end{center}
\end{table}

\section*{Acknowledgments}

The authors thank E. Roache and J. Melnick for their technical assistance.
We gratefully acknowledge support from the VERITAS collaboration and 
University of Utah. This recearch is supported by 
National Science Foundation under NSF Grant \#0079704 and by 
grants from the U.S. Department of Energy, Enterprise
Ireland and PPARC in the UK.

\section{References}

\re
1.\ Bahcall J.N. and Wolf R.A.\ 1976, ApJ 209, 214
\re
2.\ Bekki K.\ et al.\ 2001, ApJ 557, L39
\re
3.\ Bergstrom L., Ullio P.\ 1997, Nucl. Phys. B504, 27
\re 
4.\ Bergstrom L., Ullio P., \& Buckley J.\ 1998, Astroparticle Phys. 9, 137
\re
5.\ Cesarini A.\ et al.\ 2003, astro-ph/0305075 
\re
6.\ Corbelli E. \& Salucci P.\ 2000, MNRAS 311, 441
\re
7.\ Corbin M.R.\ et al.\ 2001, AJ 121, 2549
\re
8.\ Del Burgo C.\ et al.\ 2001, MNRAS 321, 227 
\re
9.\ Dubus G. \& Rutledge R.E.\ 2002, MNRAS 336, 901
\re
10.\ Eskridge P.B. \& Schweitzer A.E.\ 2001, AJ 122, 3106  
\re
11.\ Gebhardt K.\ et al.\ 2001, AJ 122, 2469
\re
12.\ Goldberg H.\ 1983, Phys. Rev. Lett. 50, 1419
\re
13.\ Gondolo P. \& Silk J.\ 1999, Phys. Rev. Lett. 83, 1719
\re
14.\ Hennawi J.F. \& Ostriker J.P.\ 2002, ApJ 572, 41
\re
15.\ Joseph C.L.\ et al.\ 2001, ApJ 550, 668 
\re
16.\ Jungman G., Kamionkowski M., \& Griest K.\ 1996, Phys. Rev. 267, 195
\re
17.\ Kleyna J.T.\ et al.\ 1998, AJ 115, 2359
\re
18.\ Klypin A.\ et al.\ 1999, ApJ 522, 82
\re
19.\ Kosak K.\ et al.\ 2003, This proceedings
\re
20.\ Landau L.D. \& Lifshitz E.M.\ 1960, in ``Mechanics'' 
(Oxford, Pergamon Press)
\re
21.\ La Parola V.\ et al.\ 2003, ApJ 583, 758
\re
22.\ Lauer T.R.\ et al.\ 1998, AJ 116, 2263
\re
23.\ Merritt D. \& Quinlan G.D.\ 1998, ApJ 498, 625
\re
24.\ Merritt D., Ferrarese L, Joseph C.L.\ 2001, Science 293, 1116 
\re
25.\ Merritt D.\ et al.\ 2002, Phys. Rev. Lett. 88, 1301
\re
26.\ Merritt D.\ 2003, astro-ph/0301365 
\re
27.\ Milosavljevic M.\ et al.\ 2002, MNRAS 331, L51 
\re
28.\ Moore B.\ et al.\ 1999, MNRAS 310, 1147
\re
29.\ Stephens A.W. \& Frogel J.A.\ 2002, AJ 124, 2023
\re
30.\ Tremaine S.\ et al.\ 1994, AJ 107, 634
\re
31.\ Tyler G.\ 2002,  Phys. Rev. D66, 3509
\re
32.\ Ullio P., Bergstrom L.\ 1998, Phys. Rev. D57, 1962

\endofpaper
\end{document}